\documentclass[prd,twocolumn,showpacs,superscriptaddress,nofootinbib,floatfix,showkeys,10pt]{revtex4-2}

\usepackage{graphicx}
\usepackage{amsmath}
\usepackage{bm}
\usepackage{yhmath}
\usepackage{mathtools}
\usepackage{wasysym}
\usepackage[colorlinks,citecolor=blue,urlcolor=blue,linkcolor=blue]{hyperref}
\usepackage{subfigure}
\usepackage{color}
\usepackage{cases}
\usepackage{subfigure}
\usepackage{times}
\usepackage{dcolumn,booktabs,bm}
\usepackage{slashed}
\usepackage{amsfonts,amssymb,stmaryrd,latexsym,amsmath}
\usepackage{textcomp}
\usepackage{multirow}
\usepackage{cancel}
\usepackage{array}
\usepackage{orcidlink}
\usepackage{enumitem}

\renewcommand{\arraystretch}{1.8}

\begin{document}


\title{Fully strange tetraquark resonant states as the cousins of $X(6900)$}

\author{Yao Ma\,\orcidlink{0000-0002-5868-1166}}\email{yaoma@pku.edu.cn}
\affiliation{School of Physics and Center of High Energy Physics, Peking University, Beijing 100871, China}

\author{Wei-Lin Wu\,\orcidlink{0009-0009-3480-8810}}\email{wlwu@pku.edu.cn}
\affiliation{School of Physics, Peking University, Beijing 100871, China}

\author{Lu Meng\,\orcidlink{0000-0001-9791-7138}}\email{lu.meng@rub.de}
\affiliation{Institut f\"ur Theoretische Physik II, Ruhr-Universit\"at Bochum,  D-44780 Bochum, Germany }

\author{Yan-Ke Chen\,\orcidlink{
0000-0002-9984-163X}}\email{chenyanke@stu.pku.edu.cn}
\affiliation{School of Physics, Peking University, Beijing 100871, China}

\author{Shi-Lin Zhu\,\orcidlink{0000-0002-4055-6906}}\email{zhusl@pku.edu.cn}
\affiliation{School of Physics and Center of High Energy Physics, Peking University, Beijing 100871, China}

\begin{abstract}

We conduct systematic calculations of the S-wave fully strange systems with ``normal" $\left(J^{P C}=0^{++}, 1^{+-}, 2^{++}\right)$ and ``exotic" $\left(J^{P C}=0^{+-}, 1^{++}, 2^{+-}\right)$ C-parities, which are the 
strange analogue of the fully charmed tetraquark state $X(6900)$.
Within a constituent quark potential model, we employ the Gaussian expansion method to solve the four-body Schr\"odinger equation and the complex scaling method to identify resonant states. We obtain a series of resonant states and zero-width states in the mass range of 2.7 to 3.3 GeV, with their widths ranging from less than 1 MeV to about 50 MeV. Their rms radii strongly indicate that they are compact tetraquark states. Among these states, the $T_{4s,2^{++}}(2714)$ may be the most likely one to be observed experimentally. We urge the experimental exploration of the $2^{++}$ $s s \bar{s} \bar{s}$ state around 2.7 GeV in the $\phi\phi$ channel. Since the lowest S-wave $s s \bar{s} \bar{s}$ state is around 2.7 GeV, the compact P-wave $s s \bar{s} \bar{s}$ states are expected to be heavier. Hence, $\phi(2170)$ and $X(2370)$ are unlikely to be compact tetraquark states.

\end{abstract}

\maketitle

\section{Introduction}~\label{sec:intro}

Since 2003, the interest in multiquark states has been reignited by the experimental discovery of a series of candidates (for recent reviews, see Refs.~\cite{Chen:2016qju,Hosaka:2016pey,Lebed:2016hpi,Esposito:2016noz,Ali:2017jda,Guo:2017jvc,Brambilla:2019esw,Liu:2019zoy,Chen:2022asf,Meng:2022ozq,Mai:2022eur}). These discoveries have significantly enriched the hadron spectrum. The study of their properties and structures enhances our understanding of the non-perturbative nature of the strong interaction.

Inspired by their many strangeness decays, there are several promising candidates for the fully strange tetraquark states. 

\begin{itemize}
 \item 
 The state $\phi(2170)$ (was $Y(2175)$) was first reported in 2006 by the BaBar Collaboration in the process of $e^{+} e^{-} \rightarrow \phi(1020) f_0(980)$~\cite{BaBar:2006gsq,BaBar:2007ptr,BaBar:2007ceh,BaBar:2011btv}. Later it was confirmed by the BES and BESIII Collaborations in the $J / \psi \rightarrow \eta \phi f_0(980)$ process~\cite{BES:2007sqy,BESIII:2014ybv,BESIII:2017qkh}, and the Belle Collaboration in the $e^{+} e^{-} \rightarrow \phi \pi^{+} \pi^{-}$and $e^{+} e^{-} \rightarrow \phi f_0(980)$ processes~\cite{Belle:2008kuo}. 
 
 \item
 The state $X(2370)$ was first observed in 2010 by the BESIII Collaboration in the $\pi^{+} \pi^{-} \eta^{\prime}$ invariant-mass spectrum~\cite{BESIII:2010gmv}. Later it was confirmed in 2019 by the BESIII Collaboration in the $K \bar{K} \eta^{\prime}$ invariant-mass distribution~\cite{BESIII:2019wkp}, which suggests that the $X(2370)$ likely contains four valence $s$ or $\bar{s}$ quarks.

 \item
 The pseudoscalar state $X(2500)$ was first observed in 2016 by the BES Collaboration in the process $J / \psi \rightarrow \gamma \phi \phi$~\cite{BESIII:2016qzq}, together with the scalar state $f_0(2100)$, and three tensor states, the $f_2(2010), f_2(2300)$ and $f_2(2340)$.

 \item
 The $X(2239)$ was seen in 2018 by the BESIII Collaboration in the cross section of $e^{+} e^{-} \rightarrow K^{+} K^{-}$~\cite{BESIII:2018ldc}.
 
 \item
 An evidence for a structure $X(2100)$ in the $\phi \eta^{\prime}$ mass spectrum of the $J / \psi \rightarrow \phi \eta \eta^{\prime}$ decay was reported in 2019 by the BESIII Collaboration~\cite{BESIII:2018zbm}.

 \item 
 Another structure in the $\phi f_0$ cross section around 2.4 GeV, labelled as $X(2436)$~\cite{Shen:2009mr}, was reported in 2007 by the BaBar Collaboration~\cite{BaBar:2007ptr}. It is also seen by the BES Collaboration~\cite{BES:2007sqy} and Belle Collaboration~\cite{Belle:2008kuo}, and evidenced by the BESIII Collaboration~\cite{BESIII:2014ybv}. However, the latest BESIII  results indicate that its statistical significance is no more than 2$\sigma$~\cite{BESIII:2021lho}.

\end{itemize}
Among the aforementioned possible $T_{s s \bar{s} \bar{s}}$ candidates, all the established ones have negative parity. The existence of the positive parity states remains uncertain. Additionally, the LHCb Collaboration recently discovered a fully charmed tetraquark candidate $X(6900)$~\cite{LHCb:2020bwg}, which was later confirmed by the CMS and ATLAS~\cite{CMS:2023owd, ATLAS:2023bft}. 
The QCD Lagrangian leads to the same quark gluon interaction if one omits the quark mass difference. The discovery of $X(6900)$ implies the possible existence of the strange analogue $T_{s s \bar{s} \bar{s}}$. 

Inspired by these experimental observations, various theoretical investigations have been conducted on the fully strange tetraquark systems. Regarding the $\phi(2170)$ state with $J^{P C}=1^{--}$, some authors used the QCD sum rule~\cite{Wang:2006ri,Chen:2008ej,Chen:2018kuu,Jiang:2023atq}, quark model~\cite{Deng:2010zzd} and diquark-antidiquark scheme~\cite{Drenska:2008gr} to interpret it as an $s s \bar{s} \bar{s}$ tetraquark state, although other interpretations were not  excluded~\cite{Ding:2006ya,Ding:2007pc,MartinezTorres:2008gy,Alvarez-Ruso:2009vkn,Dong:2017rmg,Ho:2019org,Agaev:2019coa}. Similarly, the $X(2370)$ state was also explained as an $s s \bar{s} \bar{s}$ tetraquark state with $0^{-+}$~\cite{Dong:2020okt,Su:2022eun}. The pseudoscalar state $X(2500)$ was interpreted as a $P$-wave $s s \bar{s} \bar{s}$ tetraquark state of $J^{P C}=0^{-+}$~\cite{Liu:2020lpw,Dong:2020okt,Su:2022eun}. In Ref.~\cite{Lu:2019ira}, the $X(2500)$, $X(2239)$, $f_2(2300)$ were assigned as $0^{-+}$, $1^{--}$, $2^{++}$ $s s \bar{s} \bar{s}$ tetraquark states, respectively. Moreover, the structure $X(2100)$ was interpreted as a $1^{+-}$ $s s \bar{s} \bar{s}$ state~\cite{Cui:2019roq,Wang:2019nln,Su:2022eun}. In Ref.~\cite{Azizi:2019ecm}, the resonant states $X(2100)$ and $X(2239)$ were treated as the axial-vector and vector tetraquarks with the quark content $s s \bar{s} \bar{s}$, respectively. The structure $X(2436)$ was explained as a $P$-wave $s s \bar{s} \bar{s}$ tetraquark state of $J^{P C}=1^{--}$~\cite{Jiang:2023atq,Su:2022eun}. The $s s \bar{s} \bar{s}$ tetraquark states with other quantum numbers were also expolred ~\cite{Ebert:2008id,Liu:2020lpw,Su:2022eun,Dong:2022otb,Tan:2023azs,Xi:2023byo,Wang:2024pgy}.

So far, there is no experimental conclusion regarding the existence of positive parity $s s \bar{s} \bar{s}$ states, and theoretical investigations on these states are also limited. Although some theoretical calculations on the fully strange tetraquark system in quark models have been conducted~\cite{Ebert:2008id,Lu:2019ira,Liu:2019zoy,Tan:2023azs}, they are relatively sparse and less refined compared to those on tetraquark systems containing heavy quarks. The studies in Refs.~\cite{Ebert:2008id,Lu:2019ira} focus exclusively on the diquark-antidiquark picture, obtaining only the states above the dimeson thresholds. These states would become scattering states or resonant states once the dimeson configuration is incorporated. Only the resonant states have physical significance and should be identified using nontrivial approaches. In Refs.~\cite{Liu:2019zoy,Tan:2023azs}, the $s s \bar{s} \bar{s}$ tetraquark resonant states were investigated via the real scaling method. This method detects resonant states by avoiding the cross pattern of energy levels across a parameter controlling the spatial extension of the wave function, which is usually quite rough. 

To address these issues, we will conduct systematic calculations of the positive parity S-wave fully strange system. We will employ the Gaussian expansion method (GEM)~\cite{Hiyama:2003cu} to solve the four-body Schrödinger equation within a nonrelativistic constituent quark potential model. Additionally, we will adopt the complex scaling method (CSM)~\cite{Aguilar:1971ve,Balslev:1971vb,aoyama2006complex} to identify genuine resonant states from meson-meson scattering states. In our recent works, we have incorporated both the dimeson and diquark-antidiquark configurations to carry out benchmark calculations for tetraquark bound states~\cite{Ma:2022vqf,Ma:2023int,Meng:2023jqk} and resonant states~\cite{Chen:2023syh,Wu:2024euj,Wu:2024hrv}. We will continue to include both configurations for the fully strange tetraquark system. In this study, we will investigate both the ``normal" C-parity states ($J^{PC}=0^{++}, 1^{+-}, 2^{++}$) and the ``exotic" C-parity states ($J^{PC}=0^{+-}, 1^{++}, 2^{+-}$), with the ``exotic” C-parity systems referring to those that cannot be composed of two S-wave ground quarkonia.

This paper is arranged as follows. 
In Sec.~\ref{sec:framework}, we introduce the theoretical framework, including the Hamiltonian in the constituent quark potential, the construction of the wave function, the complex scaling method, and the approach to analyzing the spatial structures.
In Sec.~\ref{sec:results}, we present the numerical results for the properties of the fully strange resonant states.
Finally, we give a brief summary and discussion in Sec.~\ref{sec:sum}.

\section{Theoretical framework}~\label{sec:framework}

\subsection{Hamiltonian}~\label{subsec:Hamiltonian}

The nonrelativistic Hamiltonian of a four-quark system reads
\begin{align}\label{eq:Hamiltonian}
H=\sum_i^4\left(m_i+\frac{\boldsymbol{p}_i^2}{2m_i}\right)+\sum_{i<j=1}^4 V_{i j}\,,
\end{align}
where $m_i$ and $\boldsymbol{p}_i$ are the mass and momentum of quark $i$.

In this study, we will employ the AL1 constituent quark potential model~\cite{Semay:1994ht,Silvestre-Brac:1996myf} containing the one-gluon-exchange (OGE) interaction and a linear confinement term.
The AL1 potential reads 
\begin{align}\label{eq:AL1}
    V_{i j}=-\frac{3}{16}&\lambda_i^c \cdot \lambda_j^c\Big(-\frac{\kappa}{r_{ij}}+\lambda r_{i j}-\Lambda\nonumber\\
    &+\frac{8\pi\kappa'}{3m_{i}m_{j}}\frac{\exp(-r_{ij}^{2}/r_{0}^{2})}{\pi^{3/2}r_{0}^{3}}\boldsymbol{s}_{i}\cdot\boldsymbol{s}_{j}
    \Big),
\end{align}
where $r_0=A\left(\frac{2m_im_j}{m_i+m_j}\right)^{-B}$. $\lambda^c$ represents the SU(3) color Gell-Mann matrix and $\boldsymbol{s}_i$ is the spin operator of quark $i$. The parameters of the potential have been determined by fitting the meson spectra in Ref.~\cite{Silvestre-Brac:1996myf} and are listed in Table~\ref{tab:paraAL1}. The calculated meson masses in AL1 model are shown in Table~\ref{tab:meson}. 

\begin{table}[htbp]
    \centering
    \caption{The parameters in the AL1 quark potential model.}
    \label{tab:paraAL1}
    \begin{tabular*}{\hsize}{@{}@{\extracolsep{\fill}}ccccccc@{}}
        \hline\hline
        $ \kappa $ &$ \lambda { [\mathrm{GeV}^{2}]}$&$ \Lambda {\rm [GeV]} $&$ \kappa^\prime $&$ m_s {\rm [GeV]}$&$ A { [\mathrm{GeV}^{B-1}]}$&$ B $\\
        \hline
        0.5069&0.1653&0.8321&1.8609&0.577&1.6553&0.2204\\
        \hline\hline
    \end{tabular*}
\end{table}


\begin{table}[htbp] 
\renewcommand{\arraystretch}{1.4}
\centering
\caption{\label{tab:meson} The theoretical masses (in MeV) and rms radii (in fm) of the $s\bar{s}$ mesons in the AL1 model, compared with the experimental results taken
from Ref.~\cite{ParticleDataGroup:2022pth}. }
\begin{tabular*}{\hsize}{@{}@{\extracolsep{\fill}}lccccc@{}}

\hline\hline
$J^{PC}$ & Meson & $m_{\mathrm{Exp.}}$  & $m_{\mathrm{Theo.}}$  & $r^{\mathrm{rms}}_{\mathrm{Theo.}}$\\
\hline 
$0^{-+}$ & $\ensuremath{\eta^{\prime}}$~\footnote{For simplicity, we assume that there is no mixing effects between the $I=0$ $\eta(n\bar{n})$ and $\eta'(s\bar{s})$.} & - & $713.5$ & $0.54$\\
 & $\ensuremath{\eta^{\prime}(2S)}$ & - & $1565.2$ & $1.17$\\
 & $\ensuremath{\eta^{\prime}(3S)}$ & - & $2140.9$ & $1.65$\\
\hline 
$1^{--}$ & $\ensuremath{\phi}$ & $1019.5$ & $1021.0$ & $0.70$\\
 & $\ensuremath{\phi(2S)}$ & $1680$ & $1695.1$ & $1.25$\\
 & $\ensuremath{\phi(3S)}$ & $2188$ & $2231.6$ & $1.70$\\

\hline \hline
\end{tabular*}
\end{table}

\subsection{Wave function construction}~\label{subsec:wavefunction}

The wave function bases of the $ss\bar{s}\bar{s}$ tetraquark states can be expressed as the direct product of spatial wave function $\phi$, spin wave function $\chi_{s}$, and color wave function $\chi_{c}$.
\begin{equation}\label{eq:Abasis}
\psi=\mathcal{A}\left(\phi \otimes \chi_{s} \otimes \chi_{c}\right),
\end{equation}
where $\mathcal{A}$ is the antisymmetrization operator, representing the exchange of identical quarks. For the $ss\bar{s}\bar{s}$ system, $\mathcal{A}=\left(1-P_{12}\right)\left(1-P_{34}\right)$, where $P_{i j}$ permutes the $i$th and $j$th (anti)quarks.

For the spatial wave function, the Gaussian expansion method (GEM)~\cite{Hiyama:2003cu} is employed. Namely, the spatial wave function is expanded using the following basis:
\begin{equation}\label{eq:basisSpace}
\phi_{n l m}(\boldsymbol{r})=\sqrt{\frac{2^{l+5 / 2}}{\Gamma\left(l+\frac{3}{2}\right) r_n^3}}\left(\frac{r}{r_n}\right)^l e^{-\frac{r^2}{r_n^2}} Y_{l m}(\hat{r}),
\end{equation}
where the $r_n$ is taken in geometric progression, $r_n=r_0 a^{n-1}$. $Y_{l m}$ is the spherical harmonics.

\begin{figure}[htbp]
  \centering
  \includegraphics[width=0.47\textwidth]{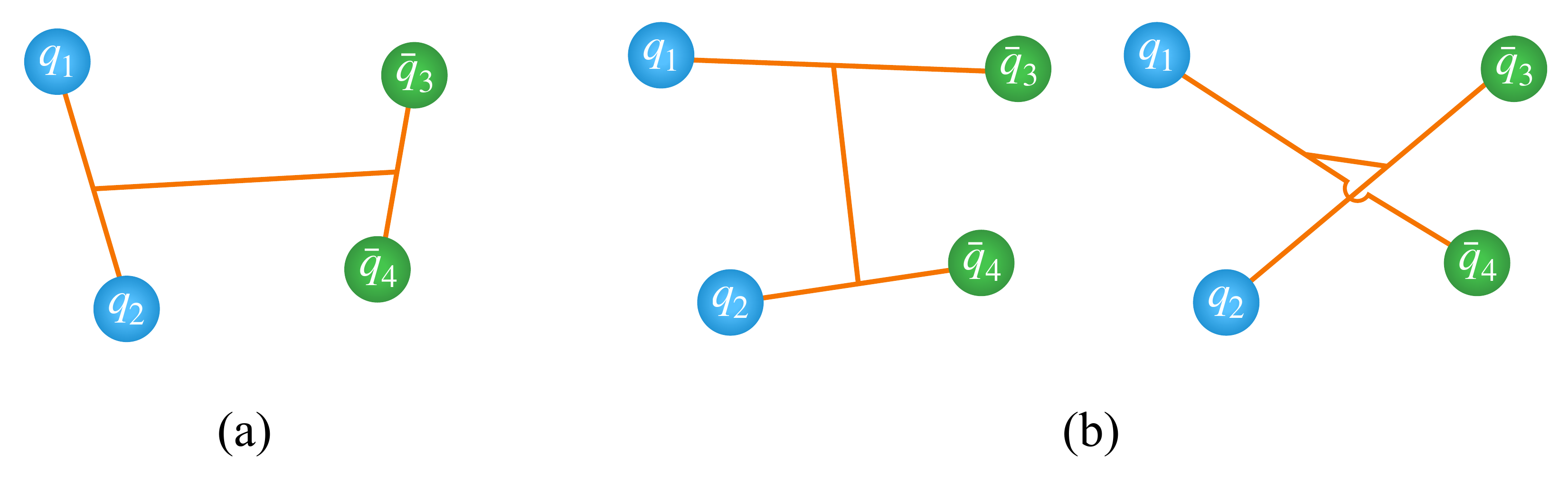} 
  \caption{\label{fig:structure} Two structures for the tetraquark system. (a) diquark-antidiquark structure. (b) dimeson structure. }
    \setlength{\belowdisplayskip}{1pt}
\end{figure}

For a four-body system, one can work on the Jacobi coordinates to exclude the center-of-mass motion. In principle, one could construct a complete basis functions using any type of Jacobi coordinates. However, to get rid of the complexity from the angular momentum, we use only S-wave bases constructed in different types of Jacobi coordinates to partially compensate for higher partial waves (see Ref.~\cite{Meng:2023jqk} for details). In this work, we include two types of Jacobi coordinates, the diquark-antidiquark structure and the dimeson structures, as shown in Fig.~\ref{fig:structure}.

In our calculation, we choose $n_{\mathrm{max}}=12$ bases for each coodinate, and the settings of the basis parameters are 
\begin{equation}
\left\{\begin{array}{ll}
r_0=0.4\ \mathrm{fm}, r_{n_{\mathrm{max}}}=2.0\ \mathrm{fm} & s-s \text { or } \bar{s}-\bar{s} \\
r_0=0.4\ \mathrm{fm}, r_{n_{\mathrm{max}}}=2.0\ \mathrm{fm} & (s s)-(\bar{s} \bar{s}) \\
r_0=0.4\ \mathrm{fm}, r_{n_{\mathrm{max}}}=1.3\ \mathrm{fm} & s-\bar{s} \\
r_0=0.5\ \mathrm{fm}, r_{n_{\mathrm{max}}}=4.5\ \mathrm{fm} & (s \bar{s})-(s \bar{s})
\end{array},\right.
\end{equation}
where we use relatively large $r_{n_{\mathrm{max}}}$ for the coordinate between two $s \bar{s}$ clusters to better describe the meson-meson scattering states and possible molecular states.

For the spin wave function, the bases for total spin $S=0,1,2$ are
\begin{equation}
\begin{aligned}
& S=0:\left\{\begin{array}{l}
{\left[\left(s_1 s_2\right)_0\left(\bar{s}_3 \bar{s}_4\right)_0\right]_0} \\
{\left[\left(s_1 s_2\right)_1\left(\bar{s}_3 \bar{s}_4\right)_1\right]_0}
\end{array},\right. \\
& S=1:\left\{\begin{array}{l}
{\left[\left(s_1 s_2\right)_0\left(\bar{s}_3 \bar{s}_4\right)_1\right]_1} \\
{\left[\left(s_1 s_2\right)_1\left(\bar{s}_3 \bar{s}_4\right)_0\right]_1} \\
{\left[\left(s_1 s_2\right)_1\left(\bar{s}_3 \bar{s}_4\right)_1\right]_1}
\end{array},\right. \\
& S=2:\begin{array}{l}
{\ \ \ \left[\left(s_1 s_2\right)_1\left(\bar{s}_3 \bar{s}_4\right)_1\right]_2}
\end{array}.
\end{aligned}
\end{equation}

Two optional color wave function bases sets are
\begin{equation}
\text { color-I: }\left\{\begin{array}{l}
{\left[\left(s_1 s_2\right)_{\bar{3}}\left(\bar{s}_3 \bar{s}_4\right)_3\right]_1} \\
{\left[\left(s_1 s_2\right)_6\left(\bar{s}_3 \bar{s}_4\right)_{\overline{6}}\right]_1}
\end{array},\right.
\end{equation}
\begin{equation}
\text { color-II: }\left\{\begin{array}{l}
{\left[\left(s_1 \bar{s}_3\right)_1\left(s_2 \bar{s}_4\right)_1\right]_1} \\
{\left[\left(s_1 \bar{s}_3\right)_8\left(s_2 \bar{s}_4\right)_8\right]_1}
\end{array}.\right.
\end{equation}
Since both sets are orthogonal and complete, using either set will yield exactly the same results~\cite{Meng:2023jqk}.

Besides the antisymmetrization operation in Eq. (\ref{eq:Abasis}), since the $s s \bar{s} \bar{s}$ system carries definite C-parity, it is necessary to add (or subtract) the C-transformation of each term in Eq. (\ref{eq:Abasis}) to ensure that the overall wave function has the desired positive (or negative) C-parity. The specific C-transformation is
\begin{align}
&\left[\left(q_1 q_2^{\prime}\right)^{s_1}_{\bar{c}}\left(\bar{q}_3 \bar{q}_4^{\prime}\right)_{c}^{s_2}\right]^S_{1_{c}}\phi\left(\boldsymbol{r}_1, \boldsymbol{r}_2, \boldsymbol{r}_3, \boldsymbol{r}_4\right)\nonumber\\
\xrightarrow{C}&\left.(-1)^{S-s_1-s_2}\left[\left(q_1 q_2^{\prime}\right)\right)_{\bar{c}}^{s_2}\left(\bar{q}_3 \bar{q}_4^{\prime}\right)_{c}^{s_1}\right]^S_{1_{c}}\phi(\boldsymbol{r}_{3},\boldsymbol{r}_{4},\boldsymbol{r}_{1},\boldsymbol{r}_{2})\nonumber
\end{align}
for the bases with color-I set, and
\begin{align}
&\left[\left(q_1 \bar{q}_3\right)_{c}^{s_1}\left(q_2^{\prime} \bar{q}_4^{\prime}\right)_{c}^{s_2}\right]^S_{1_{c}}\phi\left(\boldsymbol{r}_1, \boldsymbol{r}_2, \boldsymbol{r}_3, \boldsymbol{r}_4\right)\nonumber\\
\xrightarrow{C}&(-1)^{s_1+s_2}\left[\left(q_1 \bar{q}_3\right)_{c}^{s_1}\left(q_2^{\prime} \bar{q}_4^{\prime}\right)_{c}^{s_2}\right]^S_{1_{c}}\phi(\boldsymbol{r}_{3},\boldsymbol{r}_{4},\boldsymbol{r}_{1},\boldsymbol{r}_{2})\nonumber\\
\end{align}
for the bases with color-II set. The particle $q$ and $q'$ can be different quarks, although they both represent the same $s$ quark in the $s s \bar{s} \bar{s}$ system.

\subsection{Complex scaling method}~\label{subsec:method}

The complex scaling method (CSM) is a direct approach to obtain the energies and the decay widths of resonant states in a many-body system by performing an analytical continuation of the Schr\"odinger equation~\cite{Aguilar:1971ve,Balslev:1971vb,aoyama2006complex}. This is achieved by carrying out a complex rotation on the coordinate $\boldsymbol{r}$ and momentum $\boldsymbol{p}$, given by
\begin{align}\label{eq:complexRotation}
U(\theta) \boldsymbol{r}=\boldsymbol{r} e^{i \theta}, \quad U(\theta) \boldsymbol{p}=\boldsymbol{p} e^{-i \theta}.
\end{align}
Under the rotation, the Hamiltonian in Eq. (\ref{eq:Hamiltonian}) becomes
\begin{equation}\label{eq:HamiltonianComplex}
H(\theta)=\sum_{i=1}^4\left(m_i+\frac{p_i^2 e^{-2 i \theta}}{2 m_i}\right)+\sum_{i<j=1}^4 V_{i j}\left(r_{i j} e^{i \theta}\right).
\end{equation}
Meanwhile, for the resonant states with pole positions within the range of the rotated angle, their wave functions become normalizable by integration, thereby solvable through localized Gaussian bases in the same way as bound states. As a result, solving the complex-scaled Schr\"odinger equation will simultaneously yield the eigenenergies of bound states and resonant states within the rotated angle.

A typical pattern of the solved eigenenergies in the complex energy plane is: The bound states lie on the negative real axis of the energy plane. The continuum states align along beams originating from thresholds with $\operatorname{Arg}(E)=-2 \theta$. The resonant states with mass $M_R$ and width $\Gamma_R$ are located at $E_R=M_R-i \Gamma_R / 2$, and only those within $\left|\operatorname{Arg}\left(E_R\right)\right|<2 \theta$ can be solved. The positions of the bound and resonant states remain unchanged with the variation of the rotation angle.  One can find more details in Refs.~\cite{Lin:2022wmj,Chen:2023eri,Chen:2023syh}.

\subsection{Spatial structure}

The root-mean-square (rms) radius is a good physical quantity for reflecting the spatial structure of the tetraquark states, and a widely used criterion for distinguishing between compact and molecular tetraquark states.

The conventional definition of the rms radius under CSM is 
\begin{equation}
r_{i j}^{\mathrm{rms,C}} \equiv \operatorname{Re}\left[\sqrt{\frac{\left(\Psi(\theta)\left|r_{i j}^2 e^{2 i \theta}\right| \Psi(\theta)\right)}{\left(\Psi(\theta) \mid \Psi(\theta)\right)}}\right],
\end{equation}
where the $\Psi(\theta)$ is the obtained complex wave function of the tetraquark state. The round bra-ket represents the so-called c-product~\cite{Romo:1968tcz} defined as
\begin{equation}
\left(\phi_n \mid \phi_m\right) \equiv \int \phi_n(\boldsymbol{r}) \phi_m(\boldsymbol{r}) \mathrm{d}^3\boldsymbol{r},
\end{equation}
without taking complex conjugate of the bra-state. This procedure ensures the function inside the integral is analytic, thereby the expectation value of the physical quantity remains stable as the rotation angle changes. The rms radius calculated from the c-product is generally not real; however, its real part can still reflect the internal quark clustering behavior if the resonant state is not too broad, as discussed in Ref.~\cite{homma1997matrix}.

Nevertheless, for tetraquark states containing identical quarks, this definition may fail to identify the molecular nature of their spatial structure due to the antisymmetrization of identical quarks, as discussed in Ref.~\cite{Chen:2023syh,Wu:2024euj,Wu:2024hrv}. To eliminate this ambiguity, we have proposed a new approach to calculate the rms radii of the fully heavy $Q Q \bar{Q} \bar{Q}$ system in Ref.~\cite{Wu:2024euj}, which can also be applied to the $s s \bar{s} \bar{s}$ system. In this new approach, the total wave function is non-orthogonally decomposed into
\begin{align} 
\Psi(\theta) =  & \sum_{s_1 \leq s_2}\left[\left[\left(q_1 \bar{q}_3^{\prime}\right)_{1_c}^{s_1}\left(q_2 \bar{q}_4^{\prime}\right)_{1_c}^{s_2}\right]_{1_c}^S \phi\left(\boldsymbol{r}_1, \boldsymbol{r}_2, \boldsymbol{r}_3, \boldsymbol{r}_4;\theta\right)\right.\nonumber\\
& - \left[\left(q_2 \bar{q}_3^{\prime}\right)_{1_c}^{s_1}\left(q_1 \bar{q}_4^{\prime}\right)_{1_c}^{s_2}\right]_{1_c}^S \phi\left(\boldsymbol{r}_2, \boldsymbol{r}_1, \boldsymbol{r}_3, \boldsymbol{r}_4;\theta\right) \nonumber\\
& - \left[\left(q_1 \bar{q}_4^{\prime}\right)_{1_c}^{s_1}\left(q_2 \bar{q}_3^{\prime}\right)_{1_c}^{s_2}\right]_{1_c}^S \phi\left(\boldsymbol{r}_1, \boldsymbol{r}_2, \boldsymbol{r}_4, \boldsymbol{r}_3;\theta\right)\nonumber\\
& + \left.\left[\left(q_2 \bar{q}_4^{\prime}\right)_{1_c}^{s_1}\left(q_1 \bar{q}_3^{\prime}\right)_{1_c}^{s_2}\right]_{1_c}^S \phi\left(\boldsymbol{r}_2, \boldsymbol{r}_1, \boldsymbol{r}_4, \boldsymbol{r}_3;\theta\right)\right]\nonumber\\
= &\mathcal{A}\left[ \sum_{s_1 \leq s_2}\left[\left(q_1 \bar{q}_3^{\prime}\right)_{1_c}^{s_1}\left(q_2 \bar{q}_4^{\prime}\right)_{1_c}^{s_2}\right]_{1_c}^S \phi\left(\boldsymbol{r}_1, \boldsymbol{r}_2, \boldsymbol{r}_3, \boldsymbol{r}_4;\theta\right)\right]\nonumber\\ 
\equiv & \mathcal{A} \Psi_{13,24}(\theta),
\end{align}
where $s_1, s_2$ sum over all spin configurations with total spin $S$. The particle $q$ and $q'$ can be different quarks, although they both represent $s$ quark in the $s s \bar{s} \bar{s}$ system. We use the decomposed wave function $\Psi_{13,24}(\theta)$ to define the new rms radius
\begin{equation}
r_{ij}^{\mathrm{rms,M}} \equiv \operatorname{Re}\left[\sqrt{\frac{\left(\Psi_{13,24}(\theta)\left|r_{i j}^2 e^{2 i \theta}\right| \Psi_{13,24}(\theta)\right)}{\left(\Psi_{13,24}(\theta) \mid \Psi_{13,24}(\theta)\right)}}\right] .
\end{equation}
This newly defined rms radius can better describe the spatial structure of the molecular-type tetraquark states containing identical quarks. We will provide the results for both definitions of the rms radius in our calculation.

\section{Numerical results}~\label{sec:results}

\subsection{States with ``normal" C-parity}\label{subsec:normalC}

\begin{figure*}[htbp]
  \centering
  \includegraphics[width=1.0\textwidth]{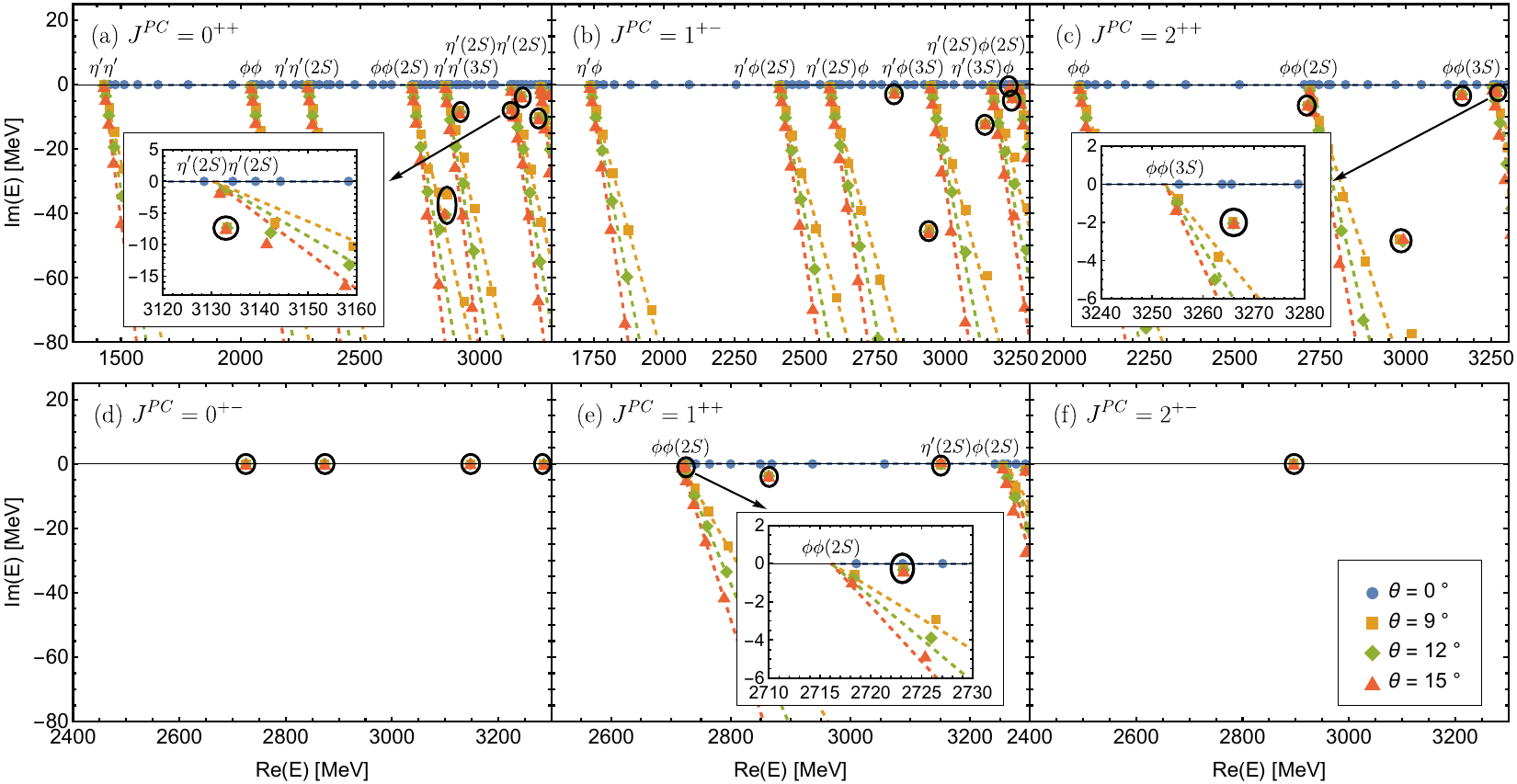} 
  \caption{\label{fig:ssss}The complex energy eigenvalues of the $ss \bar{s} \bar{s}$ states in the AL1 potential with varying $\theta$ in the CSM. The solid lines represent the continuum lines rotating along $\mathrm{Arg}(E)=-2\theta$. The resonant states do not shift as $\theta$ changes and are marked out by the black circles.}
    \setlength{\belowdisplayskip}{1pt}
\end{figure*}

The ``normal” C-parity $s s \bar{s} \bar{s}$ systems refer to those that can be composed of two S-wave ground quarkonia, namely, the  systems with $J^{PC}=0^{++},1^{+-},2^{++}$. The complex eigenenergies yielded are shown in the upper panel of Fig.~\ref{fig:ssss}. We choose different complex scaling angles $\theta=9^\circ,12^\circ,15^\circ$ to distinguish bound states and resonant states, which remain stationary as the angle changes, from the continuum states. The markers that align along beams originating from meson-meson thresholds with $\operatorname{Arg}(E) = -2 \theta$ are the continuum state eigenenergies. The results show that no bound state is found below any of the thresholds, while a series of resonant states are obtained and marked out by black circles. Their complex energies, proportions of different color configurations, and two kinds of rms radii are summarized in Table~\ref{tab:normalC}. For convenience, we label the tetraquark resonant states obtained in our calculations as $T_{4s, J^{PC}}(M)$, where $M$ is the mass of the state.

\begin{table*}[htbp] 
\renewcommand{\arraystretch}{1.4}
\centering
\caption{\label{tab:normalC} The complex energies $E=M-i \Gamma / 2$ (in MeV), proportions of different color configurations and rms radii (in fm) of the $s s \bar{s} \bar{s}$ resonant states with ``normal" C-parity in the AL1 potential. The $r_{ij}^{\mathrm{rms,C}}$ and $r_{ij}^{\mathrm{rms,M}}$ denote the conventional rms radius and the newly defined rms radius based on the decomposed wave function, respectively.
The last column shows the spatial structure of the states, where C. and M. represent the compact and molecular structure, respectively. }
\begin{tabular*}{\hsize}{@{}@{\extracolsep{\fill}}lccccccccccccc@{}}

\hline\hline
\multirow{2}{*}{$J^{PC}$} & \multirow{2}{*}{$M-i \Gamma / 2$} & \multirow{2}{*}{$\chi_{\overline{3}_c \otimes 3_c}$} & \multirow{2}{*}{$\chi_{6_c \otimes \overline{6}_c}$} & \multirow{2}{*}{$\chi_{1_{c}\otimes1_{c}}$} & \multirow{2}{*}{$\chi_{8_c \otimes 8_c}$}&\multicolumn{2}{c}{ $\ \ \ r_{ij}^{\mathrm{rms,C}}$}&\multicolumn{4}{c}{ $r_{ij}^{\mathrm{rms,M}}$}& \multirow{2}{*}{structure}\\
\cline{7-8} \cline{9-12}
&&&&&& $r_{s s}^{\mathrm{rms}}$& $r_{s \bar{s}}^{\mathrm{rms}}$& $r_{s_1 \bar{s}_3}^{\mathrm{rms}}$ & $r_{s_2 \bar{s}_4}^{\mathrm{rms}}$ & $r_{s_1 s_2}^{\mathrm{rms}}=r_{\bar{s}_3 \bar{s}_4}^{\mathrm{rms}}$ & $r_{s_1 \bar{s}_4}^{\mathrm{rms}}=r_{s_2 \bar{s}_3}^{\mathrm{rms}}$ &\\
\hline 
$0^{++}$ & $2852-40i$ & $86\%$ & $14\%$ & $38\%$ & $62\%$ & 0.95 & 1.16 & 1.20 & 1.20 & 0.91 & 1.19 &C.\\
& $2917-9i$ & $40 \%$ & $60 \%$ & $53 \%$ & $47 \%$ & 1.23 & 1.21 &  1.12 & 1.12 & 1.18 & 1.35 &C.\\
& $3133-7i$ & $58 \%$ & $42\%$ & $47\%$ & $53\%$ & 1.51 & 1.44 & 1.27 & 1.27 & 1.48 & 1.66 &C.\\
& $3175-4i$ & $46 \%$ & $54 \%$ & $51 \%$ & $49 \%$ & 1.30 & 1.27 & 1.14 & 1.14 & 1.28 & 1.39 &C.\\
& $3248-10i$ & $35 \%$ & $65 \%$ & $55 \%$ & $45 \%$ & 1.37 & 1.36 &  1.31 & 1.31 & 1.32 & 1.49 &C.\\
\hline 
$1^{+-}$ & $2819-3i$ & $63 \%$ & $37 \%$ & $46 \%$ & $54 \%$ & 1.01 & 1.11 & 1.04 & 1.05 & 1.00 & 1.18 &C.\\
& $2940-46i$ & $87 \%$ & $13 \%$ & $38 \%$ & $62 \%$ & 1.02 & 1.12 & 1.18 & 1.13 & 1.03 & 1.15 &C.\\
& $3142-12i$ & $77 \%$ & $23 \%$ & $41 \%$ & $59 \%$ & 1.15 & 1.43 & 1.51 & 1.28 & 1.10 & 1.49 &C.\\
& $3228-2i$ & $66 \%$ & $34 \%$ & $45 \%$ & $55 \%$ & 1.22 & 1.37 & 1.27 & 1.29 & 1.22 & 1.47 &C.\\
& $3237-4i$ & $64 \%$ & $36 \%$ & $45 \%$ & $55 \%$ & 1.22 & 1.33 & 1.18 & 1.22 & 1.22 & 1.44 &C.\\
\hline 
$2^{++}$ & $2714-6i$ & $75 \%$ & $25 \%$ & $42 \%$ & $58 \%$ & 1.06 & 1.09 & 1.11 & 1.11 & 0.98 & 1.14 &C.\\
& $2993-48i$ & $85 \%$ & $15 \%$ & $38 \%$ & $62 \%$ & 1.00 & 1.03 & 1.14 & 1.14 & 1.03 & 1.03 &C.\\
& $3164-3i$ & $92 \%$ & $8 \%$ & $36 \%$ & $64 \%$ & 0.94 & 1.47 &  1.47 & 1.47 & 0.93 & 1.50 &C.\\
& $3266-2i$ & $66 \%$ & $34 \%$ & $45 \%$ & $55 \%$ & 1.29 & 1.38 &  1.22 & 1.22 & 1.28 & 1.50 &C.\\

\hline \hline
\end{tabular*}
\end{table*}

The masses of the resonant states are in the region $(2.7,3.3)$ GeV, and their widths range from a few MeV to several tens of MeV. For the resonant states above the $1S2S$ di-meson threshold, our calculation actually underestimates their widths because the decay width of the $2S$ meson has not been accounted for. The rms radii of all the obtained states approximately match the sizes of the 2S mesons. Furthermore, the newly defined $r_{ij}^{\mathrm{rms,M}}$ does not exhibit any molecular characteristics, indicating that all these states are compact tetraquark states.

In tetraquark systems, there are abundant dimeson thresholds. Each of these thresholds represents a branch cut connecting two different Riemann sheets. In our calculations, we must choose appropriate rotation angle to make the pole of interest emerge. Meanwhile, due to numerical reasons, the pole close to the continuum solutions will be affected by them, as illustrated in Refs.~\cite{Wu:2024hrv,homma1997matrix}. Therefore, to reduce uncertainties, the rotation angle should be chosen such that the continuum states are not too close to the resonant state. In practices, we adjust the angle to make each pole and its rms radii stable.


Comparing the fully strange systems with $cc\bar{c}\bar{c}$ using the AL1 model in our previous work~\cite{Wu:2024euj}, we observe a similar overall pattern. However, the fully strange systems exhibit an additional quasi-bound state with $2^{++}$ below the $\phi\phi(2S)$ threshold. This difference may stem from the mass-dependent color-magnetic term in the potential.

For the $0^{++}$ states, the five resonant states can all decay into $\eta'\eta'$, $\phi\phi$, $\eta'\eta'(2S)$, $\phi\phi(2S)$ final states, although there may be some effects if the mixing between the $I=0$ $\eta(n\bar{n})$ and $\eta'(s\bar{s})$ is taken into account. The $T_{4s,0^{++}}(2917)$ and $T_{4s,0^{++}}(3133)$ have an additional decay channel into $\eta'\eta'(3S)$. The higher $T_{4s,0^{++}}(3175)$ and $T_{4s,0^{++}}(3248)$ have two additional channel into $\eta'\eta'(3S)$ and $\eta'(2S)\eta'(2S)$. For the $1^{+-}$ states, the five resonant states can all decay into $\eta'\phi$, $\eta'\phi(2S)$ and $\eta'(2S)\phi$. The  $T_{4s,1^{+-}}(3142)$ has an additional decay channel into $\eta'\phi(3S)$. And the higher $T_{4s,1^{+-}}(3228)$ and $T_{4s,1^{+-}}(3237)$ have two extra $\eta'\phi(3S)$ and $\eta'(3S)\phi$ decay channels. For the $2^{++}$ states, the lowest $T_{4s,2^{++}}(2714)$ below the $\phi\phi(2S)$ threshold have only one decay channel $\phi\phi$. Other two $2^{++}$ resonant states $T_{4s,2^{++}}(2993)$ and $T_{4s,2^{++}}(3164)$ have an additional decay channel into $\phi\phi(2S)$. The $T_{4s,2^{++}}(3266)$ has an extra decay channel to $\phi\phi(3S)$.

Among these states, the state $T_{4s,2^{++}}(2714)$ is the most likely one to be observed experimentally. Its mass is the lowest, and its decay channel $\phi\phi$ predominantly decays into $K^+K^-K^+K^-$, which are all charged particles that can be detected directly by detectors, leading to high detection efficiency and resolution.

\subsection{States with ``exotic" C-parity}\label{subsec:exoticC}

The ``exotic” C-parity systems refer to those that cannot be composed of two S-wave ground quarkonia, namely, the system with $J^{PC}=0^{+-},1^{++},2^{+-}$. The complex eigenenergies yielded are shown in the lower panel of Fig.~\ref{fig:ssss}. We obtain a series of resonant states and zero-width states, whose energies, proportions of different color configurations, and two kinds of rms radii are summarized in Table~\ref{tab:exoticC}. The rms radii approximately match the sizes of the 2S mesons. The newly defined $r_{ij}^{\mathrm{rms,M}}$ does not exhibit any molecular characteristics, indicating that all these states are compact tetraquark states. 

For the $J^{PC}=0^{+-}$ and $2^{+-}$ systems, there is no S-wave di-quarkonium threshold. So the calculated states appear as bound states on the complex energy plane. However, there may actually be higher-wave thresholds below these states, which are not included in the calculation. If the higher-wave mesons are included, the properties of these states may change and their widths may no longer be zero.

For the $J^{PC}=1^{++}$ system, the masses of the resonant states range in $(2.7,3.3)$ GeV, and their widths are small, with the largest being only a few MeV. Generally, one would expect the widths of these resonant states to arise from their decays into lower-energy meson-meson channels. However, their narrow widths suggest that such decays are suppressed. The reason for the suppression has been discussed in detail in Ref.~\cite{Wu:2024euj}. Simply put, when the rms radii between different $i,j$ quarks are roughly equal, the transition operator of the color-electric term becomes approximately proportional to the identity operator. Therefore, the color mixing transition is suppressed. In the $s s \bar{s} \bar{s}$ system, however, the contribution from the color-magnetic term is not as significantly suppressed by the quark mass as it is in the heavy quark sector. Therefore, the transition can still occur via spin mixing, which, although limited, can lead to slightly larger widths of these resonant states. All three $J^{PC}=1^{++}$ resonant states can decay via $\phi\phi(2S)$, though this process is suppressed.

\begin{table*}[htp] 
\renewcommand{\arraystretch}{1.4}
\centering
\caption{\label{tab:exoticC} The complex energies $E=M-i \Gamma / 2$ (in MeV), proportions of different color configurations and rms radii (in fm) of the $s s \bar{s} \bar{s}$ resonant states with ``exotic" C-parity in the AL1 potential. The $r_{ij}^{\mathrm{rms,C}}$ and $r_{ij}^{\mathrm{rms,M}}$ denote the conventional rms radius and the newly defined rms radius based on the decomposed wave function, respectively.
The last column shows the spatial structure of the states, where C. and M. represent the compact and molecular structure, respectively.  }
\begin{tabular*}{\hsize}{@{}@{\extracolsep{\fill}}lccccccccccccc@{}}

\hline\hline
\multirow{2}{*}{$J^{PC}$} & \multirow{2}{*}{$M-i \Gamma / 2$} & \multirow{2}{*}{$\chi_{\overline{3}_c \otimes 3_c}$} & \multirow{2}{*}{$\chi_{6_c \otimes \overline{6}_c}$} & \multirow{2}{*}{$\chi_{1_{c}\otimes1_{c}}$} & \multirow{2}{*}{$\chi_{8_c \otimes 8_c}$}&\multicolumn{2}{c}{ $\ \ \ r_{ij}^{\mathrm{rms,C}}$}&\multicolumn{4}{c}{ $r_{ij}^{\mathrm{rms,M}}$}& \multirow{2}{*}{structure}\\
\cline{7-8} \cline{9-12}
&&&&&& $r_{s s}^{\mathrm{rms}}$& $r_{s \bar{s}}^{\mathrm{rms}}$& $r_{s_1 \bar{s}_3}^{\mathrm{rms}}$ & $r_{s_2 \bar{s}_4}^{\mathrm{rms}}$ & $r_{s_1 s_2}^{\mathrm{rms}}=r_{\bar{s}_3 \bar{s}_4}^{\mathrm{rms}}$ & $r_{s_1 \bar{s}_4}^{\mathrm{rms}}=r_{s_2 \bar{s}_3}^{\mathrm{rms}}$ &\\
\hline 
$0^{+-}$ & $2725$ & $33 \%$ & $67 \%$ & $56 \%$ & $44 \%$ & 1.13 & 0.96 & 0.96 & 0.96 & 1.10 & 0.96 &C.\\

 & $2873$ & $65 \%$ & $35 \%$ & $45 \%$ & $55 \%$ & 1.17 & 1.03 & 1.02 & 1.02 & 1.13 & 1.02 &C.\\
 & $3148$ & $21 \%$ & $79 \%$ & $60 \%$ & $40 \%$ & 1.44 & 1.20  & 1.20 & 1.20 & 1.41 & 1.20 &C.\\
 & $3285$ & $78 \%$ & $22 \%$ & $41 \%$ & $59 \%$ & 1.30 & 1.28 & 1.28 & 1.28 & 1.24 & 1.28 &C.\\
\hline 
$1^{++}$ & $2723-0.5 i$ & $59 \%$ & $41 \%$ & $47 \%$ & $53 \%$ & 1.14 & 1.09 & 0.90 & 1.03 & 0.99 & 1.13 &C.\\
& $2863-4 i$ & $99 \%$ & $1 \%$ & $34 \%$ & $66 \%$ & 1.07 & 1.01 & 1.05 & 0.95 & 1.07 & 1.01 &C.\\
& $3151-0.1 i$ & $66 \%$ & $34 \%$ & $45 \%$ & $55 \%$ & 1.17 & 1.31 & 1.14 & 1.23 & 1.18 & 1.42 &C.\\
\hline 
$2^{+-}$ & $2896$ & $100 \%$ & $0 \%$ & $33 \%$ & $67 \%$ & 1.10 & 1.02 & 1.02 & 1.02 & 1.10 & 1.02 &C.\\

\hline \hline
\end{tabular*}
\end{table*}

\section{Summary and Discussion}~\label{sec:sum}

We calculate the mass spectrum of the S-wave fully strange tetraquark systems with ``normal" C-parity $J^{PC}=0^{++},1^{+-},2^{++}$ and ``exoitic" C-parity $J^{PC}=0^{+-},1^{++},2^{+-}$ using the AL1 quark potential model. We employ the GEM to solve the four-body Schr\"odinger equation, and the CSM to identify genuine resonant states from meson-meson scattering states.

We obtain a series of resonant and zero-width states in the mass region $(2.7,3,3)$ GeV, with widths ranging from less than 1 MeV to around 50 MeV. Apart from the state $T_{4s,2^{++}}(2714)$, all the others lie above the $M(1 S) M^{\prime}(2 S)$ dimeson thresholds.

We calculate two kinds of rms radius for all states, the conventional rms radius $r_{ij}^{\mathrm{rms,C}}$ and the newly defined rms radius $r_{ij}^{\mathrm{rms,M}}$ based on the decomposed wave function. For all states, both kinds of rms radii approximately match the sizes of the 2S mesons, and the newly defined $r_{ij}^{\mathrm{rms,M}}$ does not exhibit any molecular characteristics, indicating that all these states are compact tetraquark states. 

Among these states, the state $T_{4s,2^{++}}(2714)$ is the most likely to be observed experimentally due to its lowest mass and its decay into $\phi\phi$, which predominantly produces directly detectable charged $K^+K^-$ pairs, ensuring high detection efficiency and resolution.

Recalling the experimental candidates for $s s \bar{s} \bar{s}$ tetraquark states, their masses are all in range of 2.1 to 2.5 GeV, whereas the lowest S-wave resonant state in our calculations is around 2.7 GeV. This suggests the states like $\phi(2170)$ and $X(2370)$ are unlikely to be compact $s s \bar{s} \bar{s}$ tetraquark states. Since the lowest S-wave state is already as high as 2.7 GeV, it's unlikely that a P-wave compact state would be lower than the S-wave ones. However, they could still be P-wave molecular-type resonant states. In this work, a state $T_{4s,2^{++}}(2714)$ is found near the threshold of ground $\phi$ combined with a radially excited $\phi(2S)$, although it does not exhibit a molecular spatial structure. Introducing P-wave thresholds might yield molecular-type resonant states near the threshold of an S-wave meson combined with a P-wave meson, which could be lower in mass. This possibility awaits careful investigation in the future.

We emphasize that the fully charmed and fully strange tetraquark resonant states share many common characteristics. Their internal structures and underlying dynamics are nearly the same. All of them are compact states. The discovery of $X(6900)$ implies the existence of its fully strange cousins, which can be searched for in the $\phi\phi$ or $\phi \phi(2S)$ channels. While $X(6900)$ can only be produced at LHC now, the fully strange tetraquark resonances may be observed at BESIII, Belle II and LHC. The future discovery of these states shall shed light on both the charmed and strange compact tetraquark states.




\begin{acknowledgements}

We are grateful to Zi-Yang Lin, Jun-Zhang Wang and Liang-Zhen Wen for the helpful discussions. This project was supported by the National Natural Science Foundation of China (11975033 and 12070131001). This project was also funded by the Deutsche Forschungsgemeinschaft (DFG, German Research Foundation, Project ID 196253076-TRR 110). The computational resources were supported by High-performance Computing Platform of Peking University.

\end{acknowledgements}

\bibliography{Ref}

\end{document}